# Security Analysis of Tunnel Field-Effect Transistor for Low Power Hardware

Shayan Taheri[#1] and Jiann-Shiun Yuan[#2]

[#] Department of Electrical and Computer Engineering,
University of Central Florida, Orlando, FL 32816, U.S.A.

*Abstract*— **Security and energy are considered as the most important parameters for designing and building a computing system nowadays. Today's attacks target different layers of the computing system (i.e. software and hardware). Introduction of new transistor technologies to the integrated circuits design is beneficial, especially for low energy requirements. The new devices have unique features and properties that provide security advantages. However, these properties may come to the aid of an adversary to design stronger attacks. Therefore, the advantages as well as the disadvantages of these technologies need to be well studied. This paper demonstrates the area and power efficiency of the tunnel field-effect transistor (TFET) technology along with analyzing its security aspects.**

## I. INTRODUCTION

The ubiquitous connectivity among computing systems is increasing that causes significant growth in the amount of data to be processed, transmitted, and stored by these systems. This situation brings a proper environment for adversaries to exploit possible backdoors in software and/or hardware to perform malicious purposes. In this regard, security and energy are considered as the most important metrics for design of computing systems nowadays. Building a secure and low energy computing system requires multidisciplinary research across different system layers, including, application, algorithm, programming language, operating system and virtual machine, instruction set architecture, microarchitecture, register transfer level (RTL) and gate-level circuit, transistor-level circuit, transistor device, material and physics. In fact, it is very challenging to achieve a computing system immune to all the software- and hardware-based attacks with considering the strict constraints on energy, performance, functionality, area, and cost.

Recently, new transistor technologies are introduced to the very large scale integration (VLSI) design for the sake of low energy consumption, especially due to the device scaling barriers of the CMOS technology. These technologies have unique features and properties that can provide security advantages. However, they present new manufacturing errors, fault models, and reliability issues. Also, their properties might come to the aid of an adversary to design stronger attacks. Therefore, the advantages as well as the disadvantages of these technologies need to be well studied. The contributions of this paper can be stated as: demonstrating the area and power efficiency of the tunnel field-effect transistor (TFET) technology, in Section 2; and security analysis of the TFET device, in Section 3. The paper is concluded in Section 4.

## II. AREA AND POWER EFFICIENCY OF TUNNEL FIELD-EFFECT TRANSISTOR

The traditional MOSFET with the ideal 60 mV/dec sub-threshold slope is no longer qualified for the near/sub-threshold computing (NSTC) regime. Due to the fact that the MOSFET current switching process works based on the temperature-dependent injection of electrons over an energy barrier, the transition slope steepness cannot be further scaled. According to the formulas for the ON and OFF currents of the MOSFET device [1], the threshold voltage ($V_t$) may not be scaled by squared two due to its exponential relationship with the OFF current. On the other hand, the overdrive parameter (VGate-Source - $V_t$) should be sufficient enough to achieve a high ON current. These cause noticeable decelerated scaling of both threshold voltage and supply voltage. A novel inter-band tunneling transistor named, Tunnel FET has been introduced to provide steeper sub-threshold slope (i.e. smaller than 60 mV/dec) [2]. The TFET device can be described as a gated p-i-n (i.e. the hole-dominant region, the intrinsic (pure) region, and the electron-dominant region) diode that has asymmetrical doping structure and operates under reverse-bias condition. It is turned ON by tunneling at the source-channel (p-i) junction through controlling the gate voltage. Figure 1 demonstrates the TFET lateral device structure. The steeper sub-threshold slope of the TFET device helps to further down scale the supply voltage and reduce the leakage currents substantially, which makes it an excellent candidate to achieve low energy consumption for the IoT applications. The comparison between the drain- source current ($I_{DS}$) versus gate-source voltage ($V_{GS}$) curves of the n-type TFET and the n-type MOSFET is shown in Figure 2. In this simulation, the InAs homo-junction TFET model from [3] is used with having these device parameters: gate width and length of 20 nm, body thickness of 5 nm, dielectric thickness of 5 nm, source doping of 4 x 1019 $cm^3$, drain doping of 6 x 1017 $cm^3$, Si FinFET S/D doping of 1 x 1020 $cm^3$. For simulating the MOSFET behavior, the CMOS 20 nm Predictive Technology Model (PTM) - Multi Gate (MG) [4] is employed. Also, both devices are connected to the supply voltage of 0.6 V. As it can be seen from the figure, the TFET device turns ON and goes to its saturation region at a smaller value of the gate-source voltage in compare to the MOSFET device.

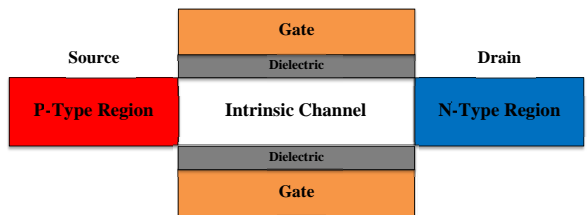

Fig. 1. TFET device lateral structure.

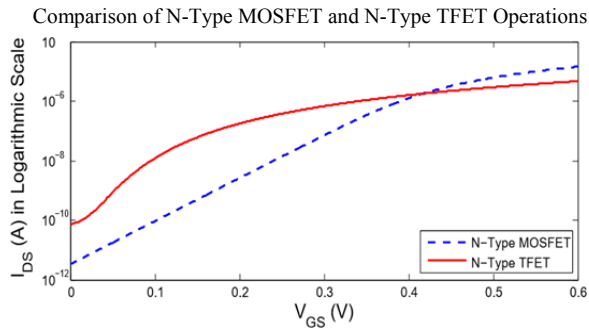

Fig. 2. The comparison between the drain-source current versus gate-source voltage curves of the n-type MOSFET and the n-type TFET.

It has been demonstrated that the TFET device has unique and unconventional features and properties that can be beneficial for all parameters of the VLSI design, such as power and security [5]. In here, the area and the average power consumption of different versions of a moderated Advanced Encryption Standard (AES) cryptographic processor [6], including its S-Box engine, a 32-Bit KATAN cryptographic hardware [7], and a canonical superscalar processing core are analyzed in both CMOS and TFET technologies. All these modules (except the processing core) are implemented in both static logic (SL) and current mode logic (CML) [8]. According to the CML style, the current bias transistor operates in its saturation region that causes providing constant current and consequently power consumption. This feature makes the CML style a good candidate for implementation of cryptographic processors to have high defense strength in front of power-based side channel attacks [9]—[11].

For the AES hardware, we can have a 128-bit data block along with 128-bit, 192-bit, or 256-bit key. This algorithm is symmetric that means the same key is used for both encryption and decryption. The primary inputs of this algorithm is the plain-text or state (i.e. a 4 x 4 column major order matrix of bytes) and the encryption/decryption key, and the primary output is the cipher-text. A plain-text is transformed to ciphertext after going through a number of transformation rounds. The typical number of transformation rounds is equal to six in addition to the key-length divided by 32. The KATAN hardware receives 32-bit and 80-bit plain-text and key respectively as its inputs. According to its structure, there are two linear feedback shift registers in parallel, with the sizes of 13-bit and 19-bit. These two registers go through two AND- and/or XOR-based computing functions for 254 iterations until the final cipher-text is generated. The selected processing core, Core-1 is generated by FabScalar tool suite [12], which is used to quickly design and generate the super-scale processors in a canonical template. The employed instruction set architecture (ISA) in this tool suite is a derivative of MIPS [13] without having load and branch delay slots. The canonical template consists of ten pipeline stages, namely, Fetch, Decode, Rename, Dispatch, Issue, Reg Read, Execute, Load/Store Unit, Writeback, and Retire.

The SPICE-level implementation of all these modules (i.e. the SPICE-based netlists) are simulated using the Synopsys CustomSim FastSPICE simulator. The duration time for transient simulation and analysis of the netlists for the cryptographic modules is 2560 ns and the average current is extracted using one data pattern for the period of 7 ns. The supply voltage is set to 0.6 V and 0.3 V in the CMOS-based and the TFET-based netlists respectively. The area of each module is calculated from the available data in the SPICE-based libraries, and its absolute value of average power consumption is obtained from the simulation data. With considering the S-Box engine as the reference module, its static logic implementation has the area and the absolute value of average power consumption of 4,484,160 ($nm^2$) and 0.2385 (pW) for CMOS technology, and 3,271,600 ($nm^2$) and 0.046371 (pW) for TFET technology. In the CML style, they are 7,470,144 ($nm^2$) and 789.78 (pW) for CMOS technology, and 5,391,880 ($nm^2$) and 47.976 (pW) for TFET technology. The proportional achieved area and average power consumption results from implementation of different versions of the AES processor are shown in Table I. All of the implemented AES processors have only one transformation round. For the super-scale processor, the duration time for transient analysis of its netlists is 2560 ns and the average current is extracted using one data pattern for the period of 500 ns (except for the TFET-based Load/Store Unit pipeline stage that is 100ns). The supply voltage is set to 0.6 V and 0.3 V in the CMOS-based and the TFET-based netlists respectively. Table II represents the analysis results for the super-scale processor. Meanwhile, the behavioral modules of performance simulation, random access memories, content addressable memories, and caches are ignored during hardware simulation. From the presented results in these two tables, it is interpreted that the TFET technology can provide less power consumption and area possession.

| | | CMOS-based SL | | CMOS-based CML | | TFET-based SL | | TFET-based CML | |
|---|---|---|---|---|---|---|---|---|---|
| | | Area | Power | Area | Power | Area | Power | Area | Power |
| 1 | KATAN S – Box | 1.1479 | 22.4046 | 0.2762 | 0.2155 | 1.0934 | 12.5229 | 0.2882 | 0.2032 |
| 2 | AES 16 – Bit / KATAN | 1.6650 | 0.2287 | 2.5042 | 2.8063 | 1.6728 | 0.2991 | 2.3635 | 2.8421 |
| 3 | AES 32 – Bit / AES 16 – Bit | 2.0016 | 2.0000 | 2.0045 | 2.0072 | 2.0016 | 1.9367 | 2.0048 | 2.0083 |
| 4 | AES 64 – Bit / AES 16 – Bit | 3.9984 | 4.0000 | 3.9955 | 3.9928 | 3.9984 | 3.8889 | 3.9952 | 3.9925 |
| 5 | AES 128 – Bit / AES 16 – Bit | 9.0048 | 9.5869 | 22.1232 | 36.7691 | 9.2912 | 9.5941 | 22.4184 | 25.7896 |

TABLE I
IMPLEMENTATION OF THE CRYPTOGRAPHIC PROCESSORS USING CMOS AND TFET TECHNOLOGIES

| Pipeline Stage | Module Name | CMOS 20 nm | | TFET 20 nm | |
|---|---|---|---|---|---|
| | | Logic Cells Area ($\mu m^2$) | Average Power Consumption ($\mu W$) | Logic Cells Area ($\mu m^2$) | Average Power Consumption ($\mu W$) |
| Fetch | FetchStage1 | 68.093568 | 131.353392 | 49.276800 | 5.5168698 |
| | Fetch1Fetch2 | 21.466944 | 16.2561738 | 15.587600 | 1.94987697 |
| | FetchStage2 | 107.395200 | 98.530692 | 77.658400 | 6.1629327 |
| | Fetch2Decode | 29.586816 | 292.787688 | 21.461600 | 9.0212262 |
| Decode | Decode | 102.881664 | 1947.47616 | 80.756800 | 46.526547 |
| | InstructionBuffer | 0.679104 | 20.4963192 | 0.558000 | 0.44968962 |
| | InstBufRename | 25.652160 | 19.3739484 | 18.626800 | 2.33282073 |
| Rename | Rename | 4.928256 | 79.675686 | 3.787200 | 2.06143578 |
| | RenameDispatch | 29.329344 | 22.2438078 | 21.297200 | 2.67418698 |
| Dispatch | Dispatch | 3.312576 | 4.4663382 | 2.399600 | 0.097017309 |
| Issue | IssueQueue | 87.561216 | 963.66756 | 63.084800 | 28.3487073 |
| | IssueqRegRead | 28.209600 | 29.7913626 | 20.482800 | 2.75909943 |
| Reg. Read | RegRead | 27.015552 | 3.72464254 | 19.574400 | 0.51231588 |
| | RegReadExecute | 29.229120 | 30.554955 | 21.222800 | 2.85142134 |
| Execute | Execute | 225.548928 | 1127.57052 | 189.281600 | 19.7540151 |
| | AgenLsu | 5.581440 | 4.19726772 | 4.052800 | 0.5046876 |
| Load/Store Unit | LoadStoreUnit | 328.352832 | 1060.89852 | 243.137200 | 38.602338 |
| Write-back | WriteBack | 14.176512 | 27.5420892 | 10.289600 | 1.64981388 |
| Retire | ActiveList | 3.656448 | 7.5163338 | 2.636000 | 0.45154854 |
| | ArchMapTable | 1.340928 | 13.8324762 | 1.092800 | 0.259969746 |

TABLE II
IMPLEMENTATION OF THE FABSCALAR PROCESSOR USING CMOS AND TFET TECHNOLOGIES

## III. SECURITY ANALYSIS OF TUNNEL FIELD-EFFECT TRANSISTOR

Even though hardware implementation of cryptographic algorithms can provide higher performance and speed, it might leak some physical information (e.g. current and delay) to be used for sensitive data extraction. Correlation power analysis (CPA) can be used as an attack on cryptographic systems to discover the relationship between the theoretical power data (extracted from the Hamming Weight model) and the actual power data (measured using the SPICE simulation). There is perfect relationship between the data variables if the correlation value is +/−1.0, and there is no correlation if it is 0. The true theoretical power signal (i.e. correctly guessed key) brings remarkable spikes (especially close to +/ −1.0) in the corresponding correlation coefficient signal. Also, if the true signal along with the false signals show remarkable spikes in a scattered form, or their average amplitudes are close to +/ − 1.0, an attacker cannot find the correct key easily. Equation 1 is used for calculation of the correlation coefficient value.

$$Correlation\ Coefficient(X,Y) = \frac{\sum(X - \mu_X).(Y - \mu_Y)}{\sqrt{\sum(X - \mu_X)^2 . \sum(Y - \mu_Y)^2}} \quad (1)$$

| 4 Number of 16-Byte Plaintexts (in Hexadecimal) | | | | | | | | | | | | | | | |
|---|---|---|---|---|---|---|---|---|---|---|---|---|---|---|---|
| 03 | F4 | 5A | 49 | 50 | DF | 5B | D1 | 22 | 1A | 0E | 23 | C9 | 85 | 10 | 39 |
| 0D | 28 | 33 | 84 | 12 | B9 | 0A | 2F | B1 | BE | D1 | 73 | 41 | D5 | DD | F9 |
| 03 | 11 | E4 | 16 | D5 | 02 | C3 | FA | C2 | 44 | 5E | 17 | 47 | 4A | 1C | EB |
| 05 | 35 | 39 | 20 | 3F | A8 | 4E | 96 | C9 | 17 | 43 | 98 | 31 | 82 | EB | B4 |
| 16 Number of 1-Byte Keys (in Hexadecimal) | | | | | | | | | | | | | | | |
| DE | 36 | 97 | F3 | 70 | 88 | 17 | 1E | E2 | 0E | 0D | 6C | 12 | 2A | F5 | C8 |

TABLE III
THE FOUR 128-BIT PLAIN-TEXTS AND SIXTEEN 8-BIT KEYS.

In this paper, the S-Box engine of the AES cryptographic processor, implemented in static logic and current mode logic styles using CMOS and TFET technologies, is used for analysis. Four 128-bit plain-texts and sixteen 8-bit keys are chosen randomly to be entered to the engine, shown in Table III. The analysis is run for 10 ns with the time step of 5 ps, and its outcome can be observed in Figures 3 and 4. As it is shown, leveraging the TFET device provides more scattered correct and false correlation coefficient signals that lead to higher hardness in finding the correct key and consequently attaining a more resilient AES processor. Also, the current mode logic style smooths the correlation coefficient signals (i.e. having very small variations) that makes the process of guessing keys even harder. However, there are still minor variations in the signals that are possible to be a source of security crack. An example is shown in Figure 5.

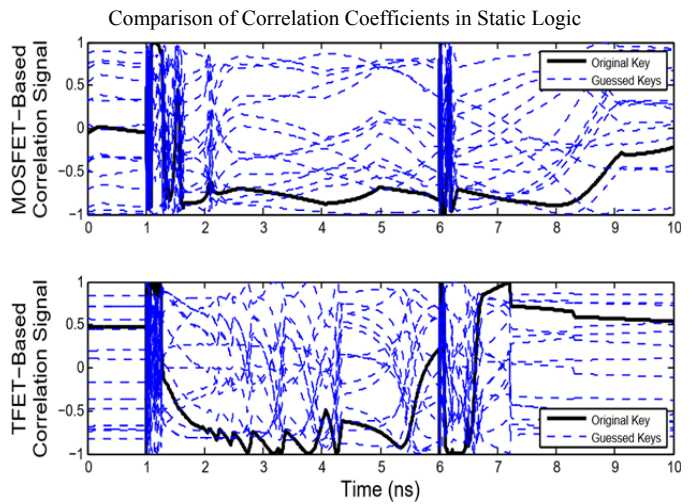

Fig. 3. The comparison of the correlation analysis on the simulation power signal and the theoretical power signal in the CMOS-based static logic style (top) and the TFET-based static logic style (bottom).

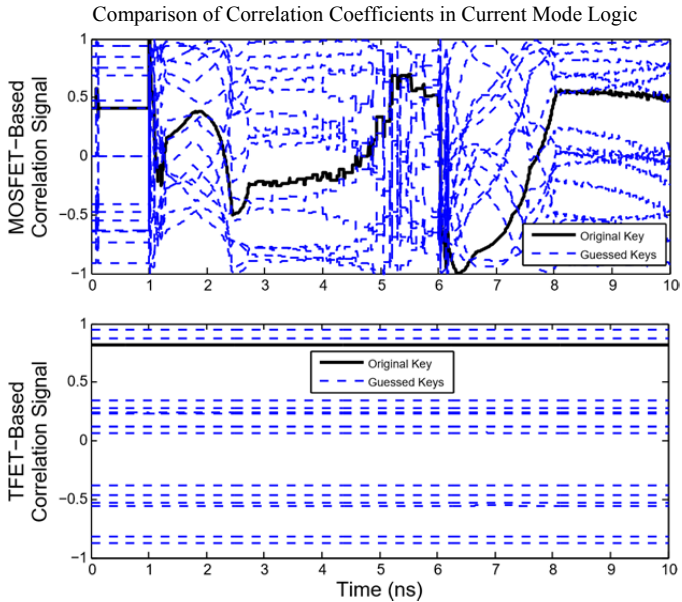

Fig. 4. The comparison of the correlation analysis on the simulation power signal and the theoretical power signal in the CMOS-based CML style (top) and the TFET-based CML style (bottom).

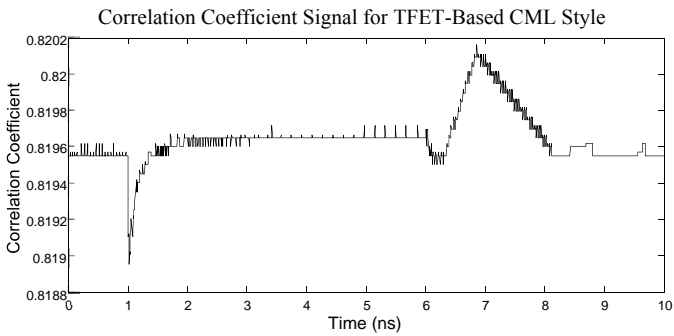

Fig. 5. An example of a correlation coefficient signal for the TFET-based CML style.

The unique features and properties of the TFET technology may have disadvantages and come to the aid of an attacker. In this regard, the footprints of two performance degradative hardware Trojans (i.e. a malicious modification to the hardware) on the area and the average power consumption of a canonical super-scale processing core are analyzed in both CMOS and TFET technologies. The FabScalar Core-1 [12] is employed for this experiment. One of the Trojans is inserted in the branch predictor module of the Fetch pipeline stage. The Fetch stage is used for selecting and translating an instruction address. The branch predictor has the duty of guessing the correct path of a branch instruction in order to improve the speed of instruction execution flow and consequently processor performance. Its prediction mechanism works based on a branch history table that includes two 2-bit saturating counters. The Trojan payload is altering these four bits and is activated according to a clock-triggered counter. The second Trojan is placed in the instruction buffer module of the Decode pipeline stage. The Decode stage interprets and decodes the selected instruction to its composition operations. The instruction buffer module reduces the instruction cache and/or main memory accesses, latency, and energy through preserving the repetitive instructions. Either reaching a counter state or a conditional state can activate the Trojan. Its payload is wrong enabling the stall signal that causes re-fetch of an instruction from the instruction cache (or the main memory).

The SPICE-level implementation of the healthy and infected modules (i.e. the SPICE-based netlists) are simulated using the Synopsys CustomSim FastSPICE simulator. The duration time for transient simulation and analysis of these netlists is 2560 ns and the average current is extracted using one data pattern for the period of 500 ns. The supply voltage is set to 0.6 V and 0.3 V in the CMOS-based and the TFET-based netlists respectively. In order to evaluate the processor performance before and after the malicious modifications, the SPEC2000 integer benchmarks are run on it for 100 million SimPoints. As it was mentioned in Section 2, the area and the absolute value of average power consumption are computed. Table IV demonstrates the analysis and evaluation results. As it can be observed from the results, the inserted Trojans degrade the processor performance noticeably. Using the TFET technology makes the Trojans to be less detectable with respect to the area occupation. It might help an attacker by causing smaller changes in the average power consumption depending on the number and type (i.e. combinational or sequential) of the logic cells used in hardware synthesis. Meanwhile, the caused overheads by the Trojans on the area and the average power consumption of the whole processor are relatively negligible.

| Module Name | Instruction Per Cycle Degradation (%) | | | | | | Logic Cells Area Change (%) | | Average Power Consumption Change (%) | |
|---|---|---|---|---|---|---|---|---|---|---|
| | bzip | gap | gzip | mcf | parser | vortex | CMOS 20nm PTM-MG | TFET 20nm InAs | CMOS 20nm PTM-MG | TFET 20nm InAs |
| Malicious Branch Prediction | 34.83 | 30.43 | 20.00 | 57.76 | 35.71 | 44.16 | 1.34 | 1.30 | 20.91 | 14.44 |
| Malicious Instruction Buffer | 77.53 | 75.36 | 72.31 | 81.03 | 78.57 | 74.03 | 1.84 | 1.82 | 17.02 | 36.92 |

TABLE IV
THE FOOTPRINTS OF TWO PERFORMANCE DEGRADATIVE HARDWARE TROJANS ON THE AREA AND THE AVERAGE POWER CONSUMPTION OF A PROCESSOR IN CMOS AND TFET TECHNOLOGIES.

IV. CONCLUSIONS

In this paper, the tunnel field-effect transistor is analyzed from area, power, and security perspectives. The area and the average power consumption of different versions of a moderated AES cryptographic processor, a 32-Bit KATAN cryptographic hardware, and a canonical super-scale processing core are calculated. The results show that the TFET device provides area and power efficiency. Two cases are studied for security analysis: (1) the S-Box engine of the AES cryptographic processor, implemented in static logic and current mode logic styles, is attacked by correlation power analysis. The TFET device coupled with the CML style brings more scattered correct and false correlation coefficient signals that leads to higher hardness in finding the correct key and consequently attaining a more resilient AES processor. (2) the footprints of two performance degradative hardware Trojans on the area and the average power consumption of a canonical superscalar processing core are analyzed. The outcome represents less detectability of the TFET-based hardware Trojans.

**Shayan Taheri** received the B.S. degree in Electrical Engineering from the Shahid Beheshti University (National University of Iran), Tehran, Iran, and the M.S. degree in Computer Engineering from the Utah State University, Logan, UT, USA, in 2013 and 2015, respectively. He is currently pursuing the Ph.D. degree in Electrical Engineering at the University of Central Florida, Orlando, FL, USA. His research interests and experiences include the applications of new transistor and memory technologies in secure and low power VLSI design, hardware Trojan design and analysis for the Internet of Things (IoT) devices, leveraging signal processing in hardware security, and VLSI Testing and Verification.

**Jiann-Shiun Yuan** received the M.S. and Ph.D. degrees from the University of Florida, Gainesville, in 1984 and 1988, respectively. In 1988 and 1989 he was with Texas Instruments Incorporated, Dallas, for CMOS DRAM design. Since 1990 he has been with the faculty of the University of Central Florida (UCF), Orlando, where he is currently a Professor and Director of NSF Multi-Functional Integrated System Technology (MIST) Center. He is the author of three textbooks and 300 papers in journals and conference proceedings. He supervised twenty-three Ph.D. dissertations, thirty-two M.S. theses, and five Honors in the Major theses at UCF. Since 1990, he has been conducting many research projects funded by the National Science Foundation, Intersil, Jabil, Honeywell, Northrop Grumman, Motorola, Harris, Lucent Technologies, National Semiconductor, and state of Florida. Dr. Yuan is a member of Eta Kappa Nu and Tau Beta Pi. He is a founding Editor of the IEEE Transactions on Device and Materials Reliability and a Distinguished Lecturer for the IEEE Electron Devices Society. He was the recipient of the 1995, 2004, 2010, and 2015 Teaching Award, UCF; the 2003 Research Award, UCF; the 2003 Outstanding Engineering Award, IEEE Orlando Section, the Excellence in Research Award at the full Professor level of the College of Engineering and Computer Science in 2015, and the Pegasus Professor Award, highest academic honor of excellence at UCF, in 2016.